# Phase space considerations for a microSAXS beamline located on a diamond Laue side-bounce monochromator

Detlef-M. Smilgies

Materials Science and Engineering, Binghamton University, Binghamton NY 13902

email: dsmilgie@binghamton.edu

**Abstract**

Flux as well as spatial and angular resolution for a microbeam small-angle x-ray scattering set-up comprising Laue optics and multiple focusing elements are modeled within five-dimensional phase space analysis. A variety of x-ray optics configurations for highest angular resolution and for highest spatial resolution are analyzed.

1. Introduction

A microbeam small-angle x-ray scattering (microSAXS) experiment has to overcome contradicting requirements: For a high resolution in SAXS the beam should have as low a divergence as possible. In order to obtain high spatial resolution the beam has to be focused to micron size, introducing enhanced divergence. The design challenge is how to achieve the desired scanning resolution while maintaining a reasonable scattering resolution at a good photon flux.

The considerations presented here are inspired by the Sector 2b and 3b side-bounce stations at the upgraded Cornell High Energy Synchrotron Source. Both stations feature a fixed scattering angle of $2\theta_B = 36°$ which corresponds to fixed beam energies of 9.7 keV, 15.9 keV, and 22.5 keV for the diamond 111, 220, and 400 reflections, respectively. Due to heatload concerns the Laue geometry is favorable as it minimizes x-ray absorption. An attractive feature of the Laue geometry is Laue focusing [1,2] which results in 1:1 focusing of the beam for symmetric cut crystals. This focusing effect has to be taken into account for properly designing the main focusing optics. In the following it is shown how phase space analysis (PSA) can be used as a convenient tool for determining flux as well as spatial and angular resolution.

## 2. Method

PSA was introduced for synchrotron radiation x-ray optics by Matsushita and Kaminaga [3,4], inspired by methods used in accelerator science. Here the variant by Pedersen and Riekel [5] will be used with extension to the full five-dimensional phase space [6]. A photon beam is considered as a distribution of rays $\rho(x,x',z,z',\eta)$ with slight deviations in position $(x,z)$, angle $(x',z')$ and energy $(\eta = \Delta E / E)$ from the optical axis. We follow here the convention that $x, x'$ are in the horizontal plane, $z, z'$ in the vertical plane and $y$ points along the beam direction. PSA relies on the paraxial approximation which works well for x-ray optics where all rays are typically close to the optical axis. By approximation of all distribution functions by Gaussians a particularly compact formulation can be found [5,6].

The x-ray flux $\Phi$ emitted by an undulator can be determined from the on-axis brilliance $B$ and the phase space density $\rho(x,x',z,z',\eta)$:

$$\Phi = B \int \rho(x,x',z,z',\eta)\, dx\, dx'\, dz\, dz'\, d\eta \tag{1}$$

The phase space density can be written in compact matrix form as

$$\rho = \exp(-0.5 X^T G_0 X) \tag{2}$$

Here $X$ is the vector of the five phase space variables, and $X^T$ its transposed. The source matrix $G_0$ for the on-axis radiation of an undulator in a straight section is diagonal and contains the inverse beam variances. A Cornell Compact Undulator [7] has a source brightness of about $5 \times 10^{17}$ photons / mm² / mrad² / 0.1% bandwidth at an electron energy of 6 GeV and a current of 200 mA. The deviations of individual rays from the optical axis are assumed to have a normal distribution with standard deviations $\sigma_x, \sigma_{x'}, \sigma_z, \sigma_{z'}, \sigma_\eta$.

Table 1. Source parameters of a Cornell Compact Undulator

| $\sigma_x$ (mm) | $\sigma_{x'}$ (mrad) | $\sigma_z$ (mm) | $\sigma_{z'}$ (mrad) | $\sigma_\eta$ (0.1%BW) |
|---|---|---|---|---|
| 0.30 | 0.087 | 0.023 | 0.013 | 26 |

As the beam propagates through the optical system, the source matrix is modified according to:

$$G_{n+1} = A_n + T_n^T G_n T_n \tag{3}$$

$A_n$ is the acceptance of the optical element and $T_n$ is the transformation of the source matrix by the optical element. The following expressions for these matrices will be relevant to our considerations:

1. Flight path of length $L$ without aperture:

$$A_{fp} = 0, \quad T_{fp}(L) = \begin{pmatrix} 1 & -L & 0 & 0 & 0 \\ 0 & 1 & 0 & 0 & 0 \\ 0 & 0 & 1 & -L & 0 \\ 0 & 0 & 0 & 1 & 0 \\ 0 & 0 & 0 & 0 & 1 \end{pmatrix}$$

$$\tag{4}$$

2. Laue Monochromator:

$$A_{mono} = \frac{1}{\sigma_D^2}\begin{pmatrix} 0 & 0 & 0 & 0 & 0 \\ 0 & 1 & 0 & 0 & -\tan(\theta_B) \\ 0 & 0 & 0 & 0 & 0 \\ 0 & 0 & 0 & 0 & 0 \\ 0 & -\tan(\theta_B) & 0 & 0 & \tan^2(\theta_B) \end{pmatrix}$$

(5)

$$T_{mono} = \begin{pmatrix} |b| & 0 & 0 & 0 & 0 \\ 0 & -1/|b| & 0 & 0 & (1-1/|b|)\tan(\theta_B) \\ 0 & 0 & 1 & 0 & 0 \\ 0 & 0 & 0 & 1 & 0 \\ 0 & 0 & 0 & 0 & 1 \end{pmatrix}$$

$\theta_B$ corresponds to the Bragg angle, $\sigma_D$ corresponds to the Darwin width of the reflection in the Gaussian approximation and *b* to the asymmetry parameter for an asymmetric cut crystal [5]. For demonstration purposes we will focus here on the case of a symmetric 220 reflection (*b*=1) with a fixed Bragg angle of 18° and a Darwin width of 0.016 in 0.1% band width yielding a photon energy of 15.9 keV.

3. Focusing element with aperture $w_{ap}$ and focal length $f$ :

$$A_{ap} = \begin{pmatrix} \sigma_{ap}^{-2} & 0 & 0 & 0 & 0 \\ 0 & 0 & 0 & 0 & 0 \\ 0 & 0 & \sigma_{ap}^{-2} & 0 & 0 \\ 0 & 0 & 0 & 0 & 0 \\ 0 & 0 & 0 & 0 & 0 \end{pmatrix} \quad T_{foc}(f) = \begin{pmatrix} 1 & 0 & 0 & 0 & 0 \\ -f^{-1} & 0 & 0 & 0 & 0 \\ 0 & 0 & 1 & 0 & 0 \\ 0 & 0 & 0 & -f^{-1} & 0 \\ 0 & 0 & 0 & 0 & 1 \end{pmatrix}$$

,

(6)

where the positional variance is given by $\sigma_{ap}^2 = w_{ap}^2 / 2\pi$.

The power of the Gaussian approximation lies therein, that instead of solving the integral in Eq.

(1), the flux can be determined directly from the transformed source matrix [6]:

$$\Phi = (2\pi)^{2.5} B \det(G)^{-0.5} \tag{7}$$

Similarly marginal standard deviations [8], for instance the standard deviation in $x$ with all other phase space variables integrated over, can be expressed by determinants [6]:

$$\sigma_{\text{marg},i} = (\det(G_{ii})/\det(G))^{0.5} \tag{8}$$

where $G_{ii}$ is the matrix obtained from $G$ by crossing out the $i^{th}$ row and $i^{th}$ column and $i$ corresponds to either $x, x', z, z'$ or $\eta$. Note that all expressions are analytical functions along the beam path $y$ and intensities and beam widths can be obtained at any point along the beam path.

If all acceptances were trivial ($A = 0$), the flux would be preserved throughout the optical system, as all *T* matrices merely shear the source density distribution which preserves the phase space volume and thus the flux. In contrast, apertures and crystal acceptance reduce the intensity, and the goal is to optimize flux hitting the sample while attaining the target beam parameters. The highest resolution obtained in small-angle scattering is both a function of beam size at the detector and the beam divergence; this makes microSAXS an interesting challenge in optics optimization [10,11].

## 3. Application

Depending on experimental requirements the beamline has to be capable of performing both high resolution SAXS and scanning microbeam SAXS [11] with either 10 µm or 1 µm resolution. Here it is assumed that the focusing elements can be removed from the beam path. Most advantageous for ease of operation are on-axis focusing elements such as x-ray compound refractive lenses or Fresnel zone plates. Figure 1 provides an overview of the three optics configurations discussed below. In the following we will focus on the medium beam energy of 15.9 keV, as obtained with a symmetric Laue 220 reflection.

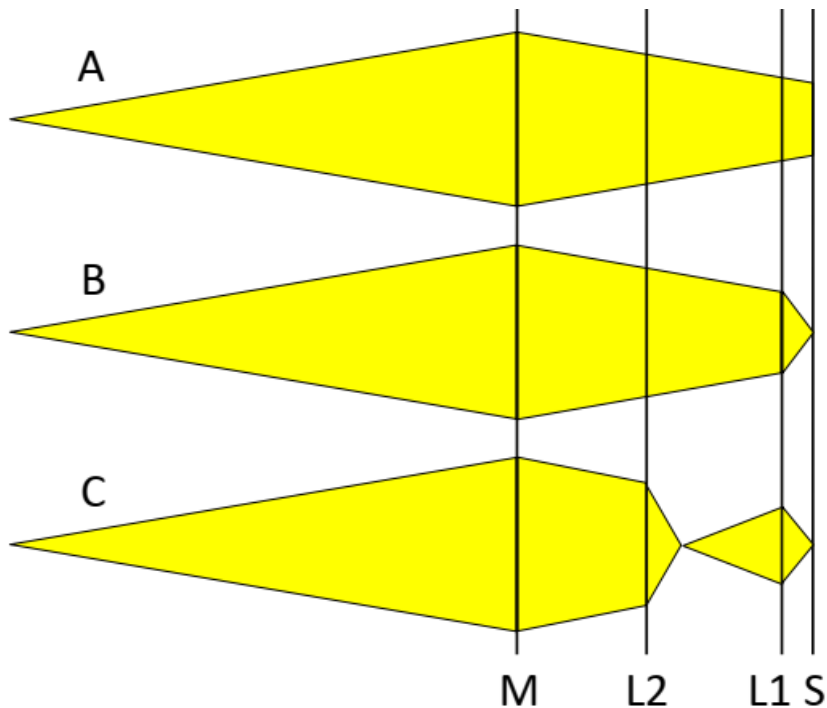

Figure 1. Focusing modes A, B, and C. M is a diamond monochromator crystal with Laue focusing, S is the sample position. L1 is the primary focusing lens. L2 is the secondary focusing lens creating a virtual source point for L1 in configuration C.

*Configuration A: Laue focusing only*

This configuration is the best to achieve high angular resolution. The beam has a 1:1 Laue focusing which preserves the divergence of the source, but reduces the beam widths. Although the detector is upstream of the 1:1 focus point, the beam diameter is reduced significantly.

The phase space description is in this case

$$G_1 = T_{fp}^T(L_{SM}) G_0 T_{fp}(L_{SM})$$
$$G_2 = A_{mono} + T_{mono}^T G_1 T_{mono}$$
$$G_3 = T_{fp}^T(L_2) G_2 T_{fp}(L_2)$$
$$G_4 = T_{fp}^T(L_{SD}) G_3 T_{fp}(L_{SD})$$

(9)

$L_{SM}$ = 16.8 m is the distance between undulator and Laue monochromator crystal which is unusually short compared with other synchrotron sources. This length is the same for all configurations; thus the beam matrices $G_1$ and $G_2$ remain the same for all three configurations. We choose the maximum sample-detector distance $L_{SD}$ = 5 m for all our considerations. Of particular interest in this configuration are the beam matrices $G_3$ and $G_4$ which contain the flux and beam widths at sample and detector, respectively. The results of the calculation are summarized in Table 1.

*Configuration B: Laue focusing and focusing element*

In configuration B we introduce another focusing element (or "lens" in short) close to the sample. $G_1$ and $G_2$ are the same as before. $L_1'$ = 7 m is now the distance between monochromator and lens, and $L_2'$ = 0.25 m the distance from lens to the sample. Compound refractive lenses with a parabolic profile have an aperture of typically $w_{ap}$ = 0.5 mm; Fresnel zone plates can even have smaller apertures.

$$\begin{aligned}
G_3' &= T_{fp}^T(L_1') G_2 T_{fp}(L_1') \\
G_4' &= A_{ap} + T_{foc}^T(f') G_3' T_{foc}(f') \\
G_5' &= T_{fp}^T(L_2') G_4' T_{fp}(L_2') \\
G_6' &= T_{fp}^T(L_{SD}) G_5' T_{fp}(L_{SD})
\end{aligned} \qquad (10)$$

As the lens is in a convergent beam due to the Laue focusing, the optical distances are now:

$$p' = L_1' - 2L_{SM}; \quad q' = L_2' \qquad (11)$$

Note that $p'$ is negative, as the virtual source point of the lens is downstream of the lens. The focusing length $f'$ is then given in the thin lens approximation as

$$f' = p'q'/(p'+q') = 0.26 \text{ m}$$

In configuration B the beam matrices $G_5'$ and $G_6'$ contain now all of the beam information at sample and detector, respectively.

*Configuration C: Laue focusing and two lenses*

In order to achieve focusing down to 1 µm, we have to use two lenses. Lens 1 remains the same as in configuration B. Lens 2 is located $L_1$" = 1.5 m from the monochromator and focuses the Laue-focused beam to the virtual source point of lens 1 at $L_2$" = 1 m. The distance from the virtual source to lens 1 is $L_3$" = 5 m. The distance of lens 1 to the sample is $L_4$"=0.25 m, as before.

$$G_3'' = T_{fp}^T(L_1'')G_2''T_{fp}(L_1'')$$
$$G_4'' = A_{ap} + T_{foc}^T(f_2'')G_3''T_{fp}(f_2'')$$
$$G_5'' = T_{fp}^T(L_2'' + L_3'')G_4''T_{fp}(L_2'' + L_3'')$$
$$G_6'' = A_{ap} + T_{foc}^T(f_1'')G_5''T_{fp}(f_1'')$$
$$G_7'' = T_{fp}^T(L_{SD})G_6''T_{fp}(L_{SD})$$
(12)

For the lens parameters we have this time:

$$\begin{array}{lll} p_2'' = L_1'' - L_{SM}, & q_2'' = L_2'', & f_2'' = p_2''q_2''/(p_2'' + q_2'') \\ p_1'' = L_3'' & q_1'' = L_4'' & f_1'' = p_1''q_1''/(p_1'' + q_1'') \end{array}$$
(13)

## 4. Results and Discussion

We are now ready calculate actual numbers. such as the flux and the the beamsize at the sample, the beamsize at the detector, and the $q$-resolution. For the maximum $q$-resolution we use the rule-of-thumb that the beamstop should be about three times as wide as the beam, so that at least 99% of the direct beam intensity is absorbed. The first resolvable peak is thus at $3\sigma_x$. This yields a minimum scattering angle $2\theta = 3\sigma_x/L_{SD}$ which corresponds to the minimum of the scattering vector:

$$q_{min} = \frac{4\pi}{\lambda}\sin(2\theta_{min}/2)$$
(14)

This in turn provides the maximum resolvable d-spacing $d_{max} = 2\pi/q_{min}$, which we will refer to as the resolution. The values for the three configurations are provided in Table 1.

Finally we will estimate the gain of the microfocusing optics. The gain is defined as the flux in the microbeam focus per unfocused flux through a pinhole of the same size as the focal spot [9].

Within the phase space analysis framework, the gain can be obtained as

$$\text{gain} = \frac{\Phi(G'_5)}{\Phi(G_3)} \frac{\sigma_{\text{marg},x}(G_3) \, \sigma_{\text{marg},z}(G_3)}{\sigma_{\text{marg},x}(G'_5) \, \sigma_{\text{marg},z}(G'_5)} \qquad (15)$$

for configuration B as well as for configuration C using $G''_7$ instead of $G'_5$.

Table 2. Flux, horizontal beam size and divergence (as standard deviations), resolution, and gain for configurations A, B, and C. For beam size, divergence, and resolution the top and bottom numbers correspond to the horizontal and vertical dimensions, respectively.

| Configuration | Flux at sample (photons/s) | beam size at sample (µm) | divergence at sample (mrad) | beam size at detector (mm) | resolution (nm) | gain |
|---|---|---|---|---|---|---|
| A | $2.4 \times 10^{12}$ | 855<br>316 | 0.087<br>0.013 | 0.47<br>0.38 | 2750<br>3420 | 1 |
| B | $2.9 \times 10^{11}$ | 7.5<br>6.2 | 0.78<br>0.75 | 3.9<br>3.2 | 335<br>402 | 700 |
| C | $1.6 \times 10^{10}$ | 1.0<br>0.3 | 1.56<br>1.41 | 3.9<br>3.8 | 334<br>345 | 5100 |

Our calculations imply that a perfect diamond Laue crystal, such as grown with the high temperature/ high pressure method, is used. Despite the narrow Darwin width of the diamond 220 reflection of $0.016 \times 10^{-3}$, such a crystal would yield a flux of $2.4 \times 10^{12}$ photons/s at 15.9 keV beam energy. A mosaic crystal would have an enhanced acceptance and yield a higher flux, however, at higher divergence and thus lower scattering resolution and a larger focal spot. Within reasonable optical parameters, the resolution for single-lens focusing in configuration B is limited to a spot size of around 8 μm. Only with two lenses (configuration C) a horizontal spot size of 1 μm can be achieved. The small vertical beamsize at the sample makes this configuration particularly interesting for grazing incidence SAXS (GISAXS). Nonetheless, the 20-times higher flux at the sample and the round beam makes configuration B an attractive option for transmission experiments requiring high intensity, but less spatial resolution. The flux losses in B and C are mostly due to the finite size of the lens apertures.

PSA offers a straightforward way to obtain beam parameters. The results are available essentially instantaneous and thus a variety of optics configurations can be simulated efficiently. The good agreement of PSA and ray tracing has been demonstrated recently [12]. We have discussed here a variety of configurations suitable for a microSAXS beamline meeting the constraints of the sector 2b and 3b beamlines. The detailed calculations for all configurations can be obtained from the author on request.

**Acknowledgements**

Tom Krawczik, Alan Pauling, and Aaron Lyndaker be thanked for sharing lay-out parameters of the beamline with me as well as Stan Stoupin for providing the Darwin widths of the diamond reflections. I would like to acknowledge Andrey Ivashov for the use of his excellent program "SMath" [13], with which all calculations were performed. Part of this work was done at the Cornell High Energy Synchrotron Source with funding via NSF award DMR-1332208.

**Disclosure**

The author declares that there are no conflicts of interest related to this article.